# Phase Diagrams of Magnetopolariton Gases

Vladimir P. Kochereshko[1,2], Mikhail V. Durnev[1,2], Lucien Besombes[3], Henri Mariette[3], Victor F. Sapega[1,2], Alexis Axitopoulos[4], Ivan G. Savenko[9,10], Timothy C.H. Liew[5], Ivan A. Shelykh[5], Alexey V. Platonov[1,2], Simeon I. Tsintzos[7], Z. Hatzopoulos[7], Pavlos Lagoudakis[4], Pavlos G. Savvidis[6,7], Christian Schneider[8], Matthias Amthor[8], Christian Metzger[8] , Martin Kamp[8], Sven Hoefling[8], Alexey Kavokin[1,3]

[1] Spin Optics Laboratory, Saint-Petersburg State University, 1, Ulianovskaya, 198504, St-Petersburg, Russia

[2] Ioffe Physical-Technical Institute, Russian Academy of Sciences, 26, Politechnicheskaya, 194021, St-Petersburg, Russia

[3] Institut Néel, CNRS/UJF 25, avenue des Martyrs - BP 166, Fr-38042 Grenoble Cedex 9, France

[4] Physics and Astronomy School, University of Southampton, Highfield, Southampton, SO171BJ, UK

[5] Division of Physics and Applied Physics, Nanyang Technological University 637371, Singapore

[6] Department of Materials Science & Technology, University of Crete, Greece

[7] IESL-FORTH, P.O. Box 1527, 71110 Heraklion, Crete, Greece

[8] Technische Physik and Wilhelm-Conrad-Röntgen-Research Center for Complex Material Systems, Universität Würzburg, D-97074 Würzburg, Am Hubland, Germany.

[9] Science Institute, University of Iceland, Dunhagi-3, IS-107, Reykjavik, Iceland

[10] Department of Applied Physics/COMP, Aalto University, PO Box 14100, 00076 Aalto, Finland

**The magnetic field effect on phase transitions in electrically neutral bosonic systems is much less studied than those in fermionic systems, such as superconducting or ferromagnetic phase transitions[1,2]. Nevertheless, composite bosons are strongly sensitive to magnetic fields: both their internal structure and motion as whole particles may be affected[3]. A joint effort of ten laboratories has been focused on studies of polariton lasers[4], where non-equilibrium Bose-Einstein condensates of bosonic quasiparticles, exciton-polaritons, may appear or disappear under an effect of applied magnetic fields. Polariton lasers based on pillar or planar microcavities were excited both optically and electrically. In all cases a pronounced dependence of the onset to lasing on the magnetic field has been observed. For the sake of comparison, photon lasing (lasing by an electron-hole plasma) in the presence of a magnetic field has been studied on the same samples as polariton lasing. The threshold to photon lasing is essentially governed by the excitonic Mott transition[5] which appears to be sensitive to magnetic fields too. All the observed experimental features are qualitatively described within a uniform model based on coupled diffusion equations for electrons, holes and excitons and the Gross-Pitaevskii equation for exciton-polariton condensates. Our research sheds more light on the physics of non-equilibrium Bose-Einstein condensates and the results manifest high potentiality of polariton lasers for spin-based quantum logic[6] applications.**

Mixed light-matter quasiparticles in the form of exciton-polaritons[7] demonstrate remarkable collective properties including polariton lasing[8], Josephson oscillations[9], vortices[10], solitons[11], optical spin Hall[12] and, possibly, spin Meissner[13] effects. Clearly, microcavities present a laboratory rich in complex many-body processes. Electrons, holes, excitons and exciton-polaritons coexist and interact giving rise to new phases or pseudo-phases[14]. Magnetic field appears to be an efficient tool of switching between some of these phases. Here we present the first detailed study of the magnetic field effect on the (out of equilibrium) phase transitions in microcavities. So far, polariton lasers are the only existing example of bosonic lasers, where coherent light is emitted spontaneously by a bosonic condensate[15]. In the meanwhile, magnetic field has been recently shown to be instrumental for the achievement of polariton lasing in electrically pumped microcavities[16,17]. Besides, it affects the second phase transition towards a photonic laser, which takes place when stimulated emission of light from electron-hole plasma starts. The switch from polariton to photon lasing[18] is associated with the exciton Mott transition[19]: the phase transition between a bosonic gas (exciton-polariton gas) and neutral plasma (electron-hole plasma). Previous experimental studies evidenced a Mott transition in semiconductor quantum wells manifested by sharp changes of the photoluminescence (PL) energy and linewidth[20,21].

This paper summarises a coordinated effort of ten laboratories aimed at understanding magnetic field induced transitions to bosonic lasing (polariton lasing) and fermionic lasing (photon lasing). A variety of state of the art structures coming from two molecular beam epitaxy facilities, including both planar and pillar microcavities with optical and electrical injection, have been studied by means of polarisation resolved magneto-photoluminescence under continuous wave (cw) and pulsed excitations.

Figure 1(a-e) shows the photoluminescence spectra of a round pillar microcavity with embedded GaAs/AlGaAs quantum wells taken at different pumping power in the presence of the external magnetic field applied along the axis of the structure (the Faraday geometry). Details of the sample structure appear in the Methods section. The structure has been excited through a micro-objective by short-duration light pulses focused to a spot with 1.5 μm diameter and having a high energy compared to the exciton transition and the cavity mode energies. These pulses create electron-hole pairs, which cool down and form excitons. The latter relax further down in energy, and eventually form a condensate of exciton polaritons, in the polariton lasing regime. At low pumping intensities (Figure 1(a)) and relatively high magnetic fields (over 7 T) we observe a narrow and intense polariton lasing mode showing a characteristic diamagnetic shift[22]. The diamagnetic shift is a signature of the exciton-polariton state: the exciton energy enhances proportionally to $B^2$. Increasing the pumping strength or lowering the field results in dramatic modifications of the spectra (presented in Figure 1(b)): the exciton-polariton mode abruptly disappears, while a strong laser line at a higher energy emerges. This line is clearly pinned to the cavity photon mode whose energy is not affected by magnetic fields. The sharp transition from the polariton to photon lasing regime is a manifestation of the abrupt excitonic Mott transition in our system. Interestingly enough, this photon laser emission line disappears again at very weak magnetic fields at the intermediate pumping power (Figure 1(c,d)).

Figure 2(a) shows how the spectra of this pillar change when modifying the pump power at a fixed magnetic field B=6 T. One can clearly distinguish the polariton and photon lasing regimes. The thresholds are identified from the peak intensity dependencies on the pumping power for the lower polariton (LP) and cavity (C) photon modes (Figure 2(b)). This analysis has been done for the full range of available pumping and magnetic field strengths which allowed the obtaining of a phase diagram shown in Figure 2(c). One can see that the threshold to polariton lasing decreases with the magnetic field increase. This tendency is confirmed by experiments done on the pillar samples of different diameters (see the Supplementary information). In contrast, the behaviour of the threshold to photon lasing is strongly non-monotonic: initially it sharply decreases, and then increases (Figure 2(c)). The switching between exciton-polariton gas and electron-hole plasma has all the features of a first order phase transition, if this term can be applied to a non-equilibrium optically driven system.

We interpret the experimental observations both for pillar and planar geometries by an interplay of the exciton Bohr radius shrinkage in the presence of the magnetic field and field-controlled diffusion of electrons, holes and excitons away from the excitation spot. The magnetic field and phase space filling both bring the exciton transition into resonance with the cavity mode, which switches on the photon lasing effect and fully suppresses exciton-polaritons in the system in the course of an avalanche Mott transition.

Our model accounts for the diffusion of electrons and holes away from the excitation spot, formation of the exciton reservoir from the electron-hole gas and formation of the polariton condensate once the density of excitons in the reservoir achieves a critical value (see the scheme in Figure 3(b)). The details of the model are summarised in the Supplementary information. The results of calculation are shown by dashed lines in the phase diagram (Figure 2(c)). The decrease of the polariton laser threshold with increase of magnetic field is due to the suppression of in-plane diffusion of electrons and holes away from the excitation spot, because the magnetic field results in an orbital movement of the carriers. This diffusion leads to the dilution of the concentration of electrons, holes and excitons at zero field. In the presence of magnetic field normal to the quantum well planes, electrons and holes cannot spread further away from the excitation area than by distances comparable with their cyclotron orbit. The suppression of in-plane diffusion of carriers has also an important impact on the excitonic Mott transition, which is governed by the condition[23]

$$\kappa n_x a_B^2 = 1$$

where $n_x$ is the in-plane concentration of excitons, $a_B$ is the exciton Bohr radius, and $\kappa$ is a coefficient depending on the geometry of the structure. The increase of $n_x$ under the pumping spot because of the magnetic field effect leads to the decrease of the threshold power for photon lasing at low magnetic fields seen in Figure 1(c). At high fields, due to the shrinkage of the exciton Bohr radius, the critical pumping power increases again.

In order to check the validity of the model and the universality of the observed effects, we have studied magneto-photoluminescence spectra of a series of planar microcavity samples (see Methods). The results of these studies are summarised in Figure 3. The samples are pumped non-resonantly by a cw laser light. The threshold to polariton lasing has been observed very clearly. It appeared to be strongly sensitive to the magnetic field (Figure 3(a)) but also to the size of the excitation spot, as the phase diagrams in the lower panels of Figure 3 show. In the case of a small (10 μm) spot, the threshold non-monotonically depends on the field. For larger spots, the initial decrease disappears, and the threshold demonstrates a steady and monotonic increase with the increase of the magnetic field. This behaviour is consistent with our kinetic model (see the scheme in Figure 3(b)). The monotonic decrease of the threshold to polariton lasing in pillar samples is replaced by a non-monotonic behaviour in planar samples and returns to a monotonic behaviour as we increase the excitation spot size (see Figure 3(d-f)). At zero field, the exciton concentration at the center of the excitation spot appears to be strongly diluted due to the diffusion of carriers if the pump spot is narrow (Figure 3(c) and Supplementary Figure

S2). On the other hand, for large spots, the diffusion has little effect on the exciton concentration, which is why the suppression of diffusion by a magnetic field does not so notably affect the threshold of polariton lasing. The increase of the threshold with a magnetic field observed for all spot sizes is due to the shrinkage of the exciton Bohr radius which leads to the reduction of the exciton radiative life-time and, consequently, emptying of the exciton reservoir.

Polariton lasers with electrical injection strongly differ from the lasers with optical pumping from the point of view of the dynamics of formation of polariton condensates. Indeed, in our samples, electron and hole injection is nearly homogeneous over the whole area of the sample, which is why the diffusion effects play little or negligible role. On the other hand, the lifetime of excitons in the structure is shortened in the presence of the magnetic field[24] which leads to the increase of the threshold for polariton lasing. Besides, the shrinkage of the exciton Bohr radius is also responsible for the increase of the threshold to photon lasing associated with the Mott transition. This tendency is experimentally confirmed for the polariton lasing (Figure 4(a)).

Figure 4(b) shows the thresholds to polariton and photon lasing at B=5 T as functions of temperature. Both thresholds decrease in full agreement with the theory, which accounts for acceleration of the acoustic phonon relaxation processes leading to electron and hole relaxation to the exciton reservoir and exciton relaxation to the polariton condensate.

To summarize, we observed magnetic-field controlled transitions to polariton and photon lasing, which can be considered as out of equilibrium Bose-Einstein condensation and Mott phase transitions. Magnetic field effects on composite bosons are of key importance for realisation of quantum logic devices based on spin properties of bosonic liquids[25], in particular, magnetic control of the Bose-Einstein condensation of exciton polaritons may be used in quantum and classical optical memories.

**Acknowledgments** We thank Jacqueline Bloch for many helpful discussions. V.K., M.D., V.F.S. and A.K. acknowledge support from the Russian Ministry of Science and Education, contract (contract No. 11.G34.31.0067). P.G.S. acknowledges support from Greek GSRT program Aristeia (grant No. 1978).
C.S., M. A. J.F., M.K and S.H. acknowledge support from the state of Bavaria.

## Methods Summary

**Sample design and fabrication**. All structures have been grown by molecular beam epitaxy. The planar sample is a high-finesse Q>16000 mircocavity formed by 5/2 λ $Al_{0.3}Ga_{0.7}As$ cavity surrounded by 32 (35) period $AlAs/Al_{0.15}Ga_{0.85}As$ top (bottom) DBR mirrors. Four sets of three 10nm GaAs quantum wells are embedded inside the cavity at the antinodes of the electric field producing a Rabi splitting of $\Omega_R$= 9.2 meV. For the micropillars studies, reactive ion etching has been applied to sculpt circular mesas with diameter ranging from 1 to 40 μm. The data shown in Figures 1,2 are taken at the positive exciton-photon detuning of about 2 meV. The data in Figure 3 are taken at the negative detuning of 7 meV. The sample for a polariton laser with electrical injection is the same as in Ref. 16.

**Experimental set-up**. The optically excited spectra were pumped by a Ti:Sa laser with pulse duration of 2 picoseconds and energy 1.62 eV, which corresponds to a range of transparency in the Bragg mirrors. Time integrated micro PL spectra of the pillar samples were recorded by a CCD detector in magnetic fields up to 11 T in the Faraday geometry with spatial resolution of about 1 μm. The planar microcavities were mounted in a gas flow sample chamber kept at T = 3 K of superconducting cryostat operating in magnetic fields up to 5 T. The PL signal was integrated in a solid angle determined by numerical aperture (NA) of the collimating lens (NA = 0.5 for 10 µm spot and NA = 0.08 for 100 µm spot). All PL experiments were performed in back scattering geometry and at about normal incidence of the light on the sample surface.

## Figures

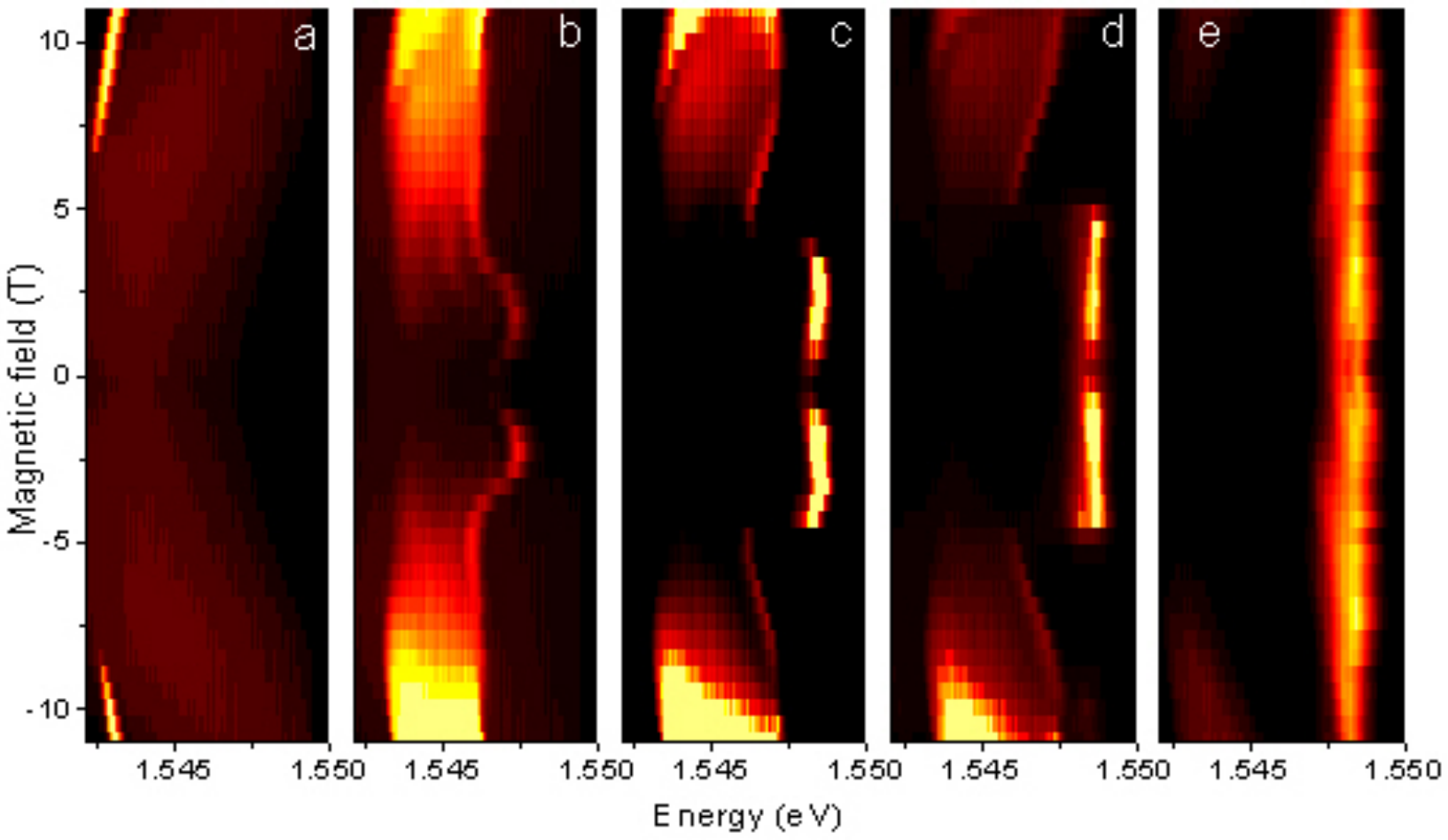

**Fig. 1. Magneto photoluminescence of the 5 $\mu$m diameter micropillar sample**. Panels represent different pumping powers: 0.09 mW (a), 0.58 mW (b), 0.85 mW (c), 1 mW (d) and 1.5 mW (e). Both polariton lasing (a) and a very sharp transition to photon lasing (c-e) are observed.

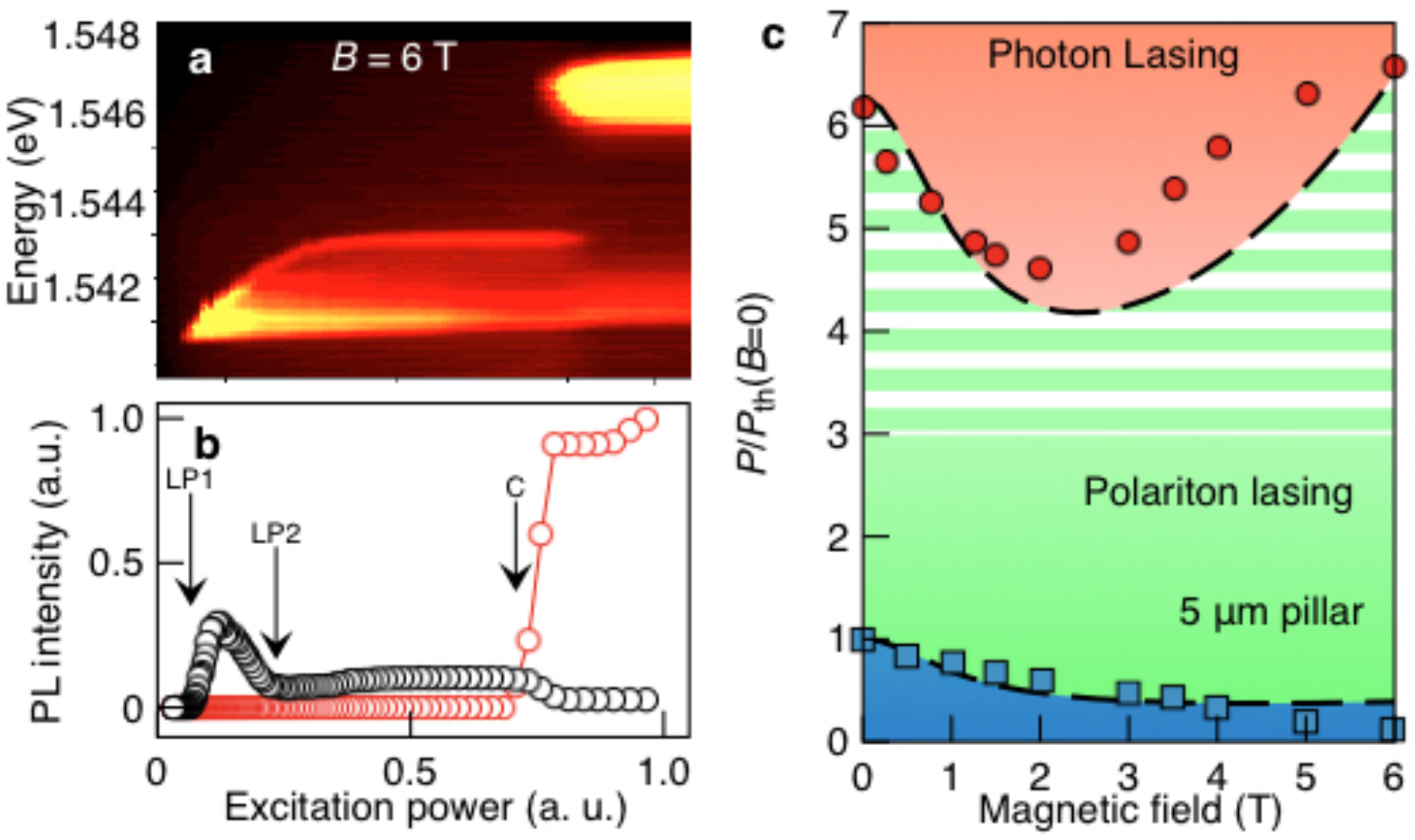

**Fig. 2. Phase transitions in micropillar samples.** a, b, Emission pattern and the integrated PL intensity of the 5 µm pillar at B = 6 T. The arrows indicate the onset and offset of the polariton laser (LP1 and LP2) and the onset of the photon laser (C). c, Phase diagram of a 5 µm pillar. Note that the polariton lasing transition at zero field corresponds to a pump power intermediate between those considered in Figure 1(a,b). The lines show the results of simulation (see Supplementary information). The part with horizontal white bands marks the polariton gas regime, beyond the offset of polariton lasing.

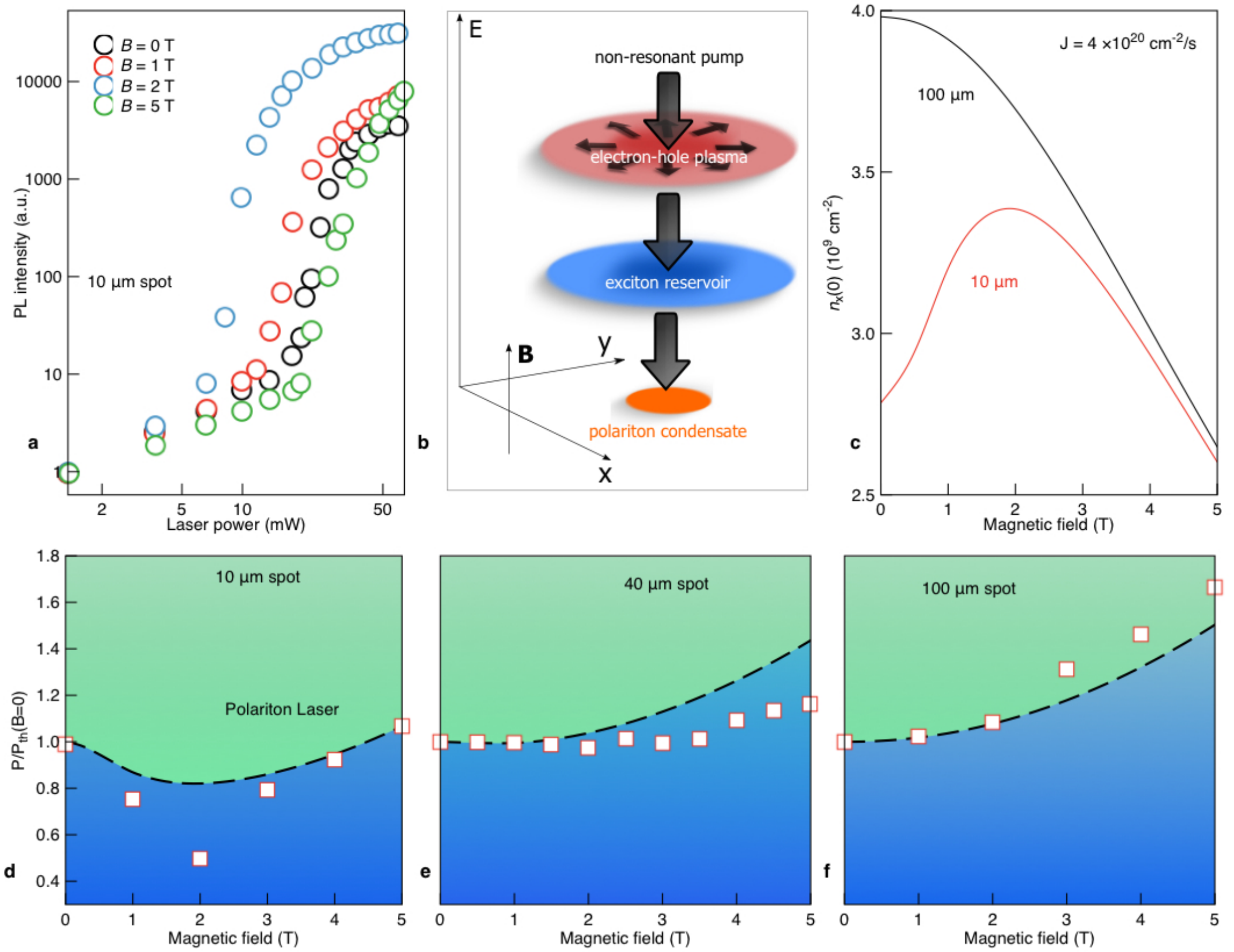

**Fig. 3. Magneto-polariton lasing of the planar microcavity sample.** a, Integrated PL intensity as a function of pump power at different magnetic fields. The laser is focused to a 10 µm spot. A clear threshold to polariton lasing is observed in the whole range of magnetic fields. b, The scheme of polariton formation under non-resonant pumping. c, Theoretical curves for the exciton density at the center of the excitation spot for a small spot (d = 10 µm, red curve) and a large spot (d = 100 µm, black curve). The dramatic difference in their behaviours reflects the crucial role of diffusion suppression by magnetic field in the case of small spots. d-f, Phae diagrams of polariton emission for three different excitation spot sizes (d = 10, 40 and 100 µm). Red squares represent the experimental data, and the results of modeling are shown by dashed curves.

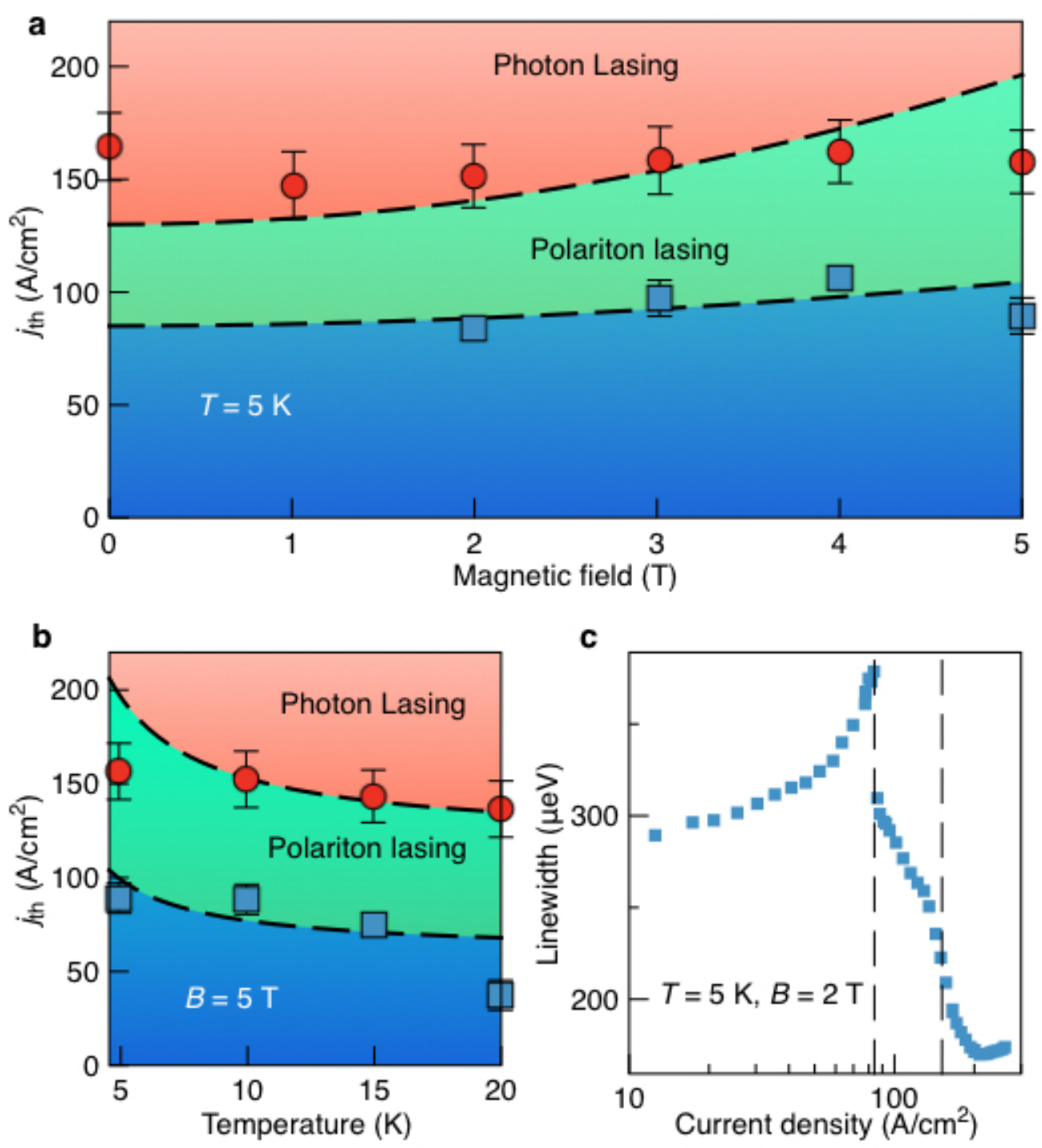

**Fig. 4. Phase diagrams of the electrically pumped polariton laser.** a,b, Magnetic and temperature dependent phase diagrams. Dashed lines show the results of our kinetic modelling. c, Linewidth of the PL emission as a function of injection current. Dashed lines indicate the onset of the polariton and photon lasing.

Semiconductor Quantum Well Microcavity: The Strong Coupling Threshold, *Phys. Rev. B* **54**, 1975-1981 (1996).

[25] Liew, T. C. H., Kavokin, A. V., and Shelykh, I. A. Optical circuits based on polariton neurons in semiconductor microcavities. *Phys. Rev. Lett.* **101**, 016402 (2008).

## SUPPLEMENTARY MATERIAL

### Basic equations

The coherent polariton state within our model is formed during the three-stage process illustrated schematically in Fig. 3b of the main text. The first stage is the nonresonant pumping of electrons and holes with large values of wave vectors; the second stage is the energy relaxation of charged carriers followed by the formation of an exciton reservoir, and the final stage is the formation of polaritons followed by its resonant scattering to the ground state. The first two processes can be described by the following set of dynamical diffusion equations:

$$\begin{aligned}\frac{\partial}{\partial t}n_e &= D_e \Delta n_e - w n_e n_h + J_e \,; \\ \frac{\partial}{\partial t}n_h &= D_h \Delta n_h - w n_e n_h + J_h \,; \\ \frac{\partial}{\partial t}n_x &= D_x \Delta n_x + w n_e n_h - \frac{n_x}{\tau_x} \,. \end{aligned} \qquad (1)$$

Here $n_e$, $n_h$ and $n_x$ are the densities of electrons, holes and excitons, respectively; $D_i$ are the diffusion coefficients, $w$ describes the exciton formation rate; $J_e$ and $J_h$ are the pump rates for electrons and holes and $\tau_x$ is the exciton lifetime. The formation of excitons in our model is described by the term $wn_en_h$ corresponding to the non-geminate (or bimolecular) mechanism because the contribution from the geminate mechanism is typically small [1, 2]. Since in a strong coupling regime non-radiative losses are negligible, we assume that excitons decay mainly due to radiative processes, so that $\tau_x$ corresponds to the radiative lifetime.

To describe the formation of a condensate the third equation in (1) should be supplemented with the scattering terms:

$$\frac{\partial n_x}{\partial t} = D_x \Delta n_x + w n_e n_h - \frac{n_x}{\tau_x} - \Gamma^{in}(|\psi_+|^2 + |\psi_-|^2)n_x. \qquad (2)$$

Here $\Gamma^{in}$ is the transition rate from the reservoir to the condensate. The equations for the condensate wavefunctions $\psi_\pm$ read:

$$i\hbar\frac{\partial \psi_\pm}{\partial t} + \frac{\hbar^2\nabla^2}{2m}\psi_\pm = -\frac{i\hbar}{2\tau}\psi_\pm + (\alpha_1|\psi_\pm|^2 + \alpha_2|\psi_\mp|^2 \pm \frac{\Delta_Z}{2})\psi_\pm + \frac{i\hbar}{2}\Gamma^{in} n_x \psi_\pm. \qquad (3)$$

Here $m$ is the effective mass of a polariton at zero wave vector, $\tau$ is a lifetime of the condensate, $\Delta_Z$ is the Zeeman splitting and $\alpha_{1,2}$ are polariton-polariton interaction constants for the parallel and anti-parallel spin configurations. The equation for the condensate wavefunction is based on a mean-field treatment [3], modified for spin [4] and incoherent excitation [5]. We note that in the present equations we neglect the drift terms, arising from the different spatial distribution of electrons and holes due to the possible difference in the diffusion coefficients $D_{e,h}$. Comparison with more accurate simulations (including the drift currents) showed the same behaviors for the phase diagrams with only minor quantitative differences, therefore we will further use $D_e = D_h \equiv D$ and omit the possibility for electrical currents inside the structure.

### Low and high pumping regimes

Let us, first, focus on the set of equations (1) for the exciton reservoir assuming the onset of polariton lasing at $n_x = n_x^{(th)}$, where $n_x^{(th)}$ is a critical exciton concentration corresponding to the population of polariton ground state equal to unity. In the following we assume that the electron and hole components are pumped with equal rates by the Gaussian-shaped laser beam $J_e = J_h \equiv J\mathrm{e}^{-r^2/R^2}$. The absence of the drift currents means equal concentrations for electrons and holes, $n_e = n_h \equiv n$, satisfying the following equation in the stationary limit:

$$D\Delta n - wn^2 + J(r) = 0\,. \tag{4}$$

Nonlinear Eq. (4) can be solved numerically, however a simple variational procedure with the carrier density taken in the form $n(r) = n_0\mathrm{e}^{-r^2/a^2}$ gives a good approximation preserving all the physical insight. Moreover, there are two limiting cases when Eq. (4) allows for analytical solutions. The first one corresponds to a low pumping regime: in that case one deals with low concentrations and the nonlinear term in Eq. (4) is small. The effective lifetime of a particle is then $\tau = 1/(wn_0)$, and the spread of carrier density is $a = \sqrt{D\tau}$. Integration of Eq. (4) over space domain gives: $a = D/(R\sqrt{wJ})$. This effective radius obviously should be greater than the radius of the pump spot, $R$, so that $a \gg R$ and $J \ll J^* = 2D^2/(wR^4)$. The excitonic density at $r = 0$ is $n_x(0) = wn_0^2\tau_x$ (if one totally neglects the diffusion of excitons, i.e. sets $D_x \equiv 0$) and it behaves quadratically with the pump intensity:

$$n_x^{\mathrm{lowpump}}(0) \propto \frac{wR^4}{2D^2}J^2\tau_x = \frac{J^2}{J^*}\tau_x\,. \tag{5}$$

In the opposite case of a high pump regime, we can neglect the diffusion term in Eq. (4) resulting in a very short linear in pump expression for the exciton density:

$$n_x^{\mathrm{highpump}}(0) = J\tau_x\,. \tag{6}$$

Since the transition between the discussed regimes is defined by $J^*$ which is inversely proportional to the fourth power of $R$, one should expect that the excitonic density distribution dependence on the pump power with and without a magnetic field significantly depends on the excitation spot radius. This effect is illustrated in Fig. S1a, where $n_x(0)$ is plotted for two spot radii different by one order of magnitude. One can see that for the sharper spot (red curve) $J^*$ significantly increases. Thus, for the fixed exciton density we switch between the high-pumping regime (black curve) and the low-pumping regime (red curve).

Application of a magnetic field alters parameters $D$ and $\tau_x$. Magnetic field leads to suppression of the diffusion of charged carriers which can be described using the Einstein relation and the Hall expression for conductivity:

$$D(B) = \frac{k_B T \mu_0 / e}{1 + B^2/B_0^2}, \quad B_0 = \frac{c}{\mu_0} \,. \tag{7}$$

Here $\mu_0 = e\tau_p/m$ is the mobility of electron/hole gas at $B = 0$ and $T$ is its temperature. The experimental data show that $T$ might be significantly higher than the lattice temperature (e.g. $T \approx 50$ K). Excitonic lifetime decreases with the magnetic field due to the decrease of the exciton Bohr radius ($\tau_x \propto a_B^2$ in 2D structures [6], see Fig. S1b). A reasonable dependence should be the Lorentzian:

$$\tau_x(B) = \frac{\tau_0}{1 + B^2/B_1^2} \tag{8}$$

with $B_1$ sufficiently larger than $B_0$ [7].

It becomes obvious now, that for the two regimes discussed above one should expect completely different behavior of excitons' density in the center of the spot as a function of magnetic field. Indeed, in the high pumping regime, $n_x(0) \propto \tau_x(B)$ decreases monotonously as a function of magnetic field, while in the low pumping regime, $n_x(0) \propto \tau_x(B)/D^2(B)$ which (in the case $B_1 > B_0$) results in the enhancement of exciton formation at $r = 0$. Further, since parameter $J^*$ depends on the magnetic field via the diffusion coefficient, the magnetic field acts as a switch between the two regimes leading to a non-monotonous behavior of the excitons' density. This is illustrated in Fig. 3c in the main text of the Letter.

**Results of calculations and parameter sets used**

As it has been already mentioned, the onset of polariton lasing is described by a condition $n_x = n_x^{(th)}$, where $n_x^{(th)}$ can be set manually. The photon lasing onset coincides with the Mott transition in the system and it is described by the following condition [8]:

$$\kappa n_x a_B^2 = 1 \,, \tag{9}$$

where $\kappa$ is the coefficient of the order of unity. In the simulations we set $\kappa = 1$. Using those two conditions, one can find the threshold pump intensity as a function of magnetic field. This function is plotted in Fig. 2c, Fig. 2d and Fig. 3d-f in the main text.

The experimental values for $w$ found in literature are: $w = 6 \pm 2$ cm$^2$/s [9], $w = 15$ cm$^2$/s [1], $w \geq 0.5$ cm$^2$/s [10]. The theoretical calculations (accounting for relaxation on optical and acoustic phonons [2]) result in $w = 0.1 \div 10$ cm$^2$/s.

The following sets of parameters were employed to calculate the phase diagrams shown in the main text:

1. **Fig. 2c**
   $D_0 = 1000$ cm$^2$/s, $B_0 = 1$ T, $B_1 = 9$ T, $w = 1$ cm$^2$/s, $n_{th} = 5 \times 10^{10}$ cm$^{-2}$, $a_B(B = 0) = 13$ nm;

2. **Fig. 2d**
   $D_0 = 1500$ cm$^2$/s, $B_0 = 2.5$ T, $B_1 = 10$ T, $w = 0.1$ cm$^2$/s, $n_{th} = 5 \times 10^{10}$ cm$^{-2}$;

3. **Fig. 3d-f**
   $D_0 = 900$ cm$^2$/s, $B_0 = 1$ T, $B_1 = 7$ T, $w = 1$ cm$^2$/s, $n_{th} = 3 \times 10^9$ cm$^{-2}$.

**Account for the coupling between the excitonic reservoir and the exciton-polariton condensate**

To investigate the role of the magnetic field on the formation of the exciton-polariton condensate from the reservoir of excitons, we used numerical analysis of the coupled Eqs. (1), (2), (3). The results of modeling of the condensate real-space profile, $|\psi|^2$, are presented in Fig. S2. One can see that, in agreement with the calculations for the excitonic reservoir, a magnetic field leads to localization of the condense density. The calculated phase diagrams describing polariton lasing show the following behavior: monotonous threshold increase in the case of the wide laser spot and non-monotonous behavior in the case of the sharp focusing. Thus, it is important to mention, that this accurate model validates the simple calculations presented above which account for the evolution of the excitonic component only.

## Equations for the laser with current injection

The situation is quite different if we consider an electrically pumped laser. In this case an electron and a hole are excited in different spots of the sample, thus leading to small values of the parameter $w$. We will assume that in our sample the electrical injection is homogeneous along the quantum well planes meaning that the diffusion is neglected in the case of current injection. On the other hand, there are lots of possible mechanisms of electron and hole decays not leading to the exciton formation (including the 'fly-through' of the particles, non-radiative processes at the pillar surface, etc.), so that the non-radiative decay times $\tau_e$ and $\tau_h$ should be introduced. The kinetic equations similar to Eq. (1)

therefore read:

$$\begin{aligned}\frac{\partial}{\partial t}n_e &= -wn_en_h - \frac{n_e}{\tau_e} + J_e \; ; \\ \frac{\partial}{\partial t}n_h &= -wn_en_h - \frac{n_h}{\tau_h} + J_h \; ; \\ \frac{\partial}{\partial t}n_x &= wn_en_h - \frac{n_x}{\tau_x} \; . \end{aligned} \tag{10}$$

In the stationary limit, we obtain the algebraic system of equations, the solution of which reads:

$$n_e = -\frac{1}{2w\tau_h} - \frac{1}{2}(J_h - J_e)\tau_e + \sqrt{\frac{1}{4}\left[\frac{1}{w\tau_h} + (J_h - J_e)\tau_e\right]^2 + J_e\frac{\tau_e}{w\tau_h}} \; , \tag{11a}$$

$$n_h = n_e\frac{\tau_h}{\tau_2} + (J_h - J_e)\tau_h \; . \tag{11b}$$

We will further make some simplifications, namely: $\tau_e = \tau_h \equiv \tau$ and $J_e = J_h \equiv J$, so that $n_e = n_h \equiv n$. In a realistic limit of small carrier lifetimes, $Jw\tau^2 < 1$, the carrier concentration is $n = J\tau$, and for the exciton density one obtains:

$$n_x = wJ^2\tau^2\tau_x \; . \tag{12}$$

We will assume a usual dependence of $\tau_x$ on magnetic field [see Eq. (8)], which gives the following expressions for the polariton and photon thresholds:

$$J_{th}^{(pol)}(B) = J_{th,0}\sqrt{1 + \frac{B^2}{B_1^2}} \; , \tag{13a}$$

$$J_{th}^{(ph)}(B) = \frac{J_{th,0}}{\sqrt{n_{th}a_B^2(0)}}\left(1 + \frac{B^2}{B_1^2}\right) \; . \tag{13b}$$

Here $J_{th,0} = \sqrt{n_{th}/(w\tau_x(0)\tau^2)}$ is the polariton threshold at zero magnetic field. The calculated phase diagram is presented in Fig. 4a in the main text with $J_{th,0} = 85$ A/cm$^2$, $B_1 = 7$ T and $n_{th}a_B^2(0) = 0.65$.

The temperature dependence of lasing thresholds is ruled by the strongly temperature-dependent exciton formation rate, $w$. We assume that the exciton formation rate in the low-temperature region increases with temperature, following: $w(T) = w_0 \exp(-T_0/T)$, where $T_0$ is some effective temperature. It leads to a decrease of the thresholds which is illustrated in Fig. 4b of the main text. The value of $T_0$ is chosen $T_0 = 5$ K.

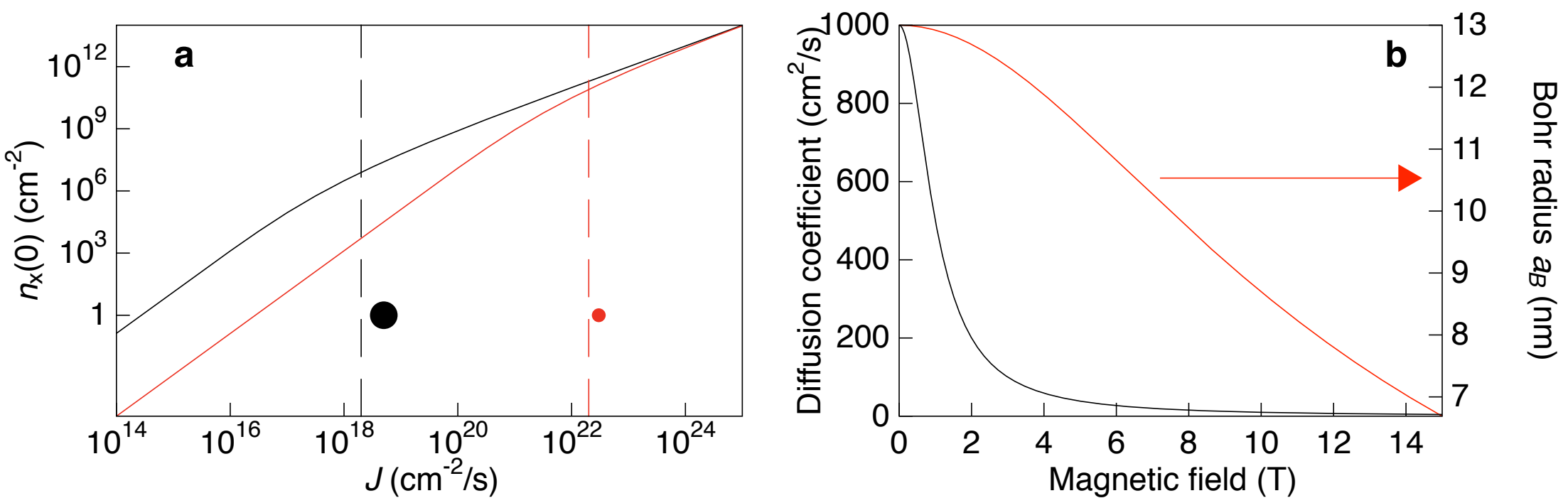

Fig. S 1: **Switch between the linear and nonlinear regimes.** a, Excitonic density in the center of the spot as a function of pump intensity. Red curve corresponds to a sharper focusing, the radii are different by one order of magnitude. The dashed lines indicate the position of $J^*$. b, Bohr radius and diffusion coefficient dependences on magnetic field as used in calculations presented in Fig. 2c in the main text.

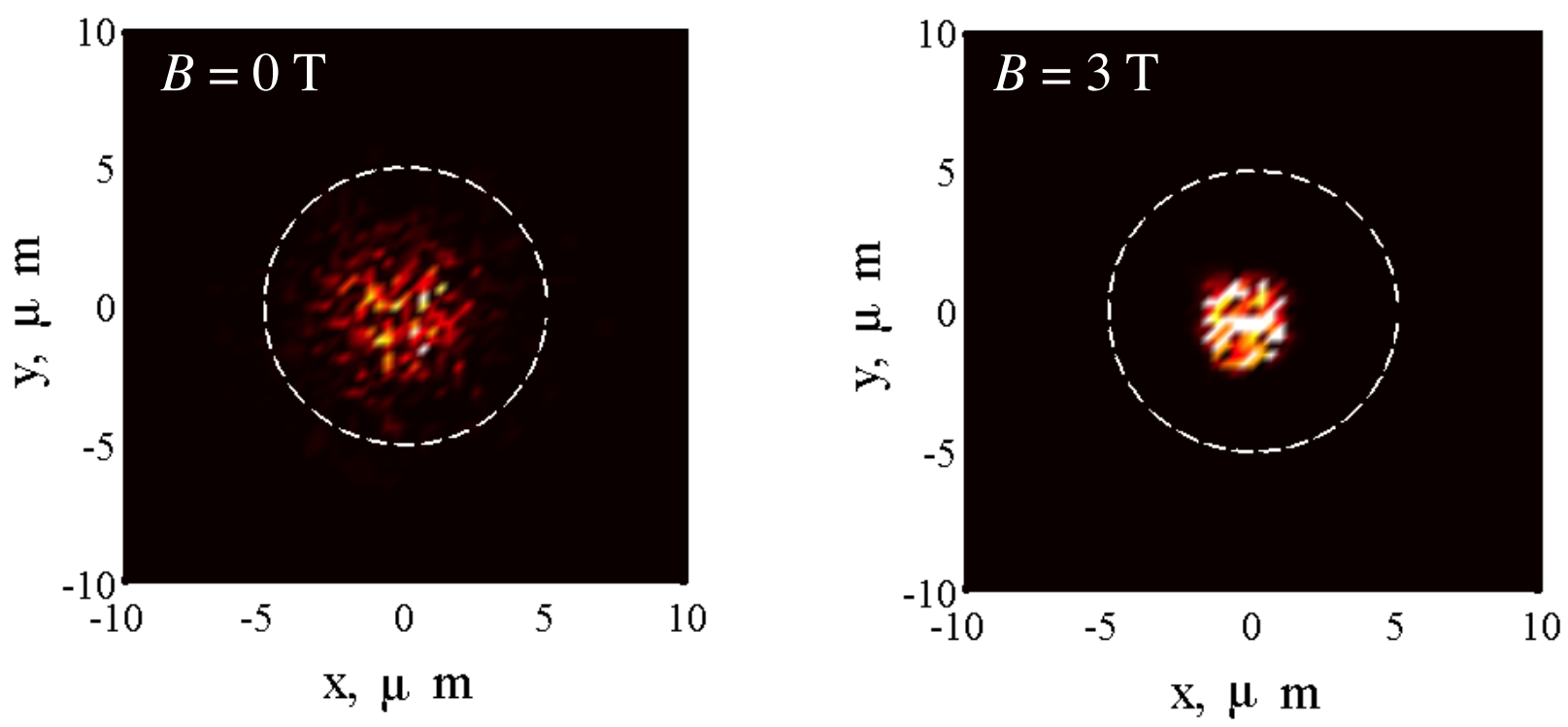

Fig. S 2: **Real-space profiles of the condensate density $|\psi(x, y)|^2$.** One can see, that application of a magnetic field leads to a stronger localization of the condensate. The excitation spot diameter is 1.5 $\mu$m. The other parameters used are: $\Gamma_{in} \approx 0.005$ $\mu$eV, $w = 1$ cm$^2$/s, $a_B = 10$ nm, $\tau_x = 100$ ps, $\tau = 10/(1 + B^2/B_1^2)$ ps, $B_0 = 1$ T, $B_1 = 7$ T, $D_0 = 1000$ cm$^2$/s, $D_x = 100$ cm$^2$/s.

**Supplementary experimental data**

This section summarizes additional experimental data not presented in the main text of the Letter.

Let us start with the micropillar samples. Figure S3 presents the data on the 8 $\mu$m diameter pillar manifesting the role of magnetic field on the in-plane motion of the carriers: due to diffusion quenching the center region of the pillar is depleted with carriers at high magnetic fields which results in lasing suppression (indicated by the arrow). On the other hand, the emission intensity increases monotonously when the pillar is excited at the center. Fig. S4 presents the experimental data on thresholds to polariton lasing measured for two micropillar samples with diameters 8 and 14 $\mu$m (the data on a 5 $\mu$m sample is presented in the main text). Both pillars clearly illustrate the suppression of diffusion by magnetic field and show the threshold behavior which correlates well with the theory. However the initial rise of the threshold in 14 $\mu$m sample is still unclear and needs further investigation.

The additional data on the planar samples is summarized in Figs. S5 and S6. Fig. S5 shows PL spectra of the sample with a planar microcavity for varying pump intensities at a fixed magnetic field $B = 5$ T and negative detuning $\delta = -7$ meV. One can see that the onset of a polariton lasing is clearly observed for both excitation spots. We conjecture that pronounced PL peaks observed at high energy side of low polaritonic branch for 10 $\mu$m spot and above threshold power originate from the cavity modes characterized by large $k$-vectors. Indeed, the lens with a larger NA (10 $\mu$m spot) integrates the PL signal in a wider $k$-space than that with a smaller NA (100 $\mu$m spot). Our additional study demonstrates that the phase diagrams (see Fig. 3d-f in the main text) are also sensitive to the detuning. Figure S6 shows the phase diagrams obtained for large 100 $\mu$m (upper panel) and small 10 $\mu$m (lower panel) spots measured at optimal negative detuning $\delta = -7$ meV (the highest efficiency of lasing) and $\delta = -3.3$ meV. One can see that deviation of the detuning from the optimal to less negative values significantly decreases the effect of magnetic field on threshold for large spot. In contrast the effect of detuning variation for small spot is almost negligible.

Figure S7 presents the Input-output characteristics and the emission linewidth of the electrically pumped microcavity sample. The polariton laser threshold is manifested by sharp increase in the intensity at 2 T and 4 T as shown in Fig. S7a, while the weak coupling laser threshold is characterized by a smooth S-curve at higher current values. The onset of polariton lasing is also accompanied by a sharp drop in the linewidth (see Fig. S7b).

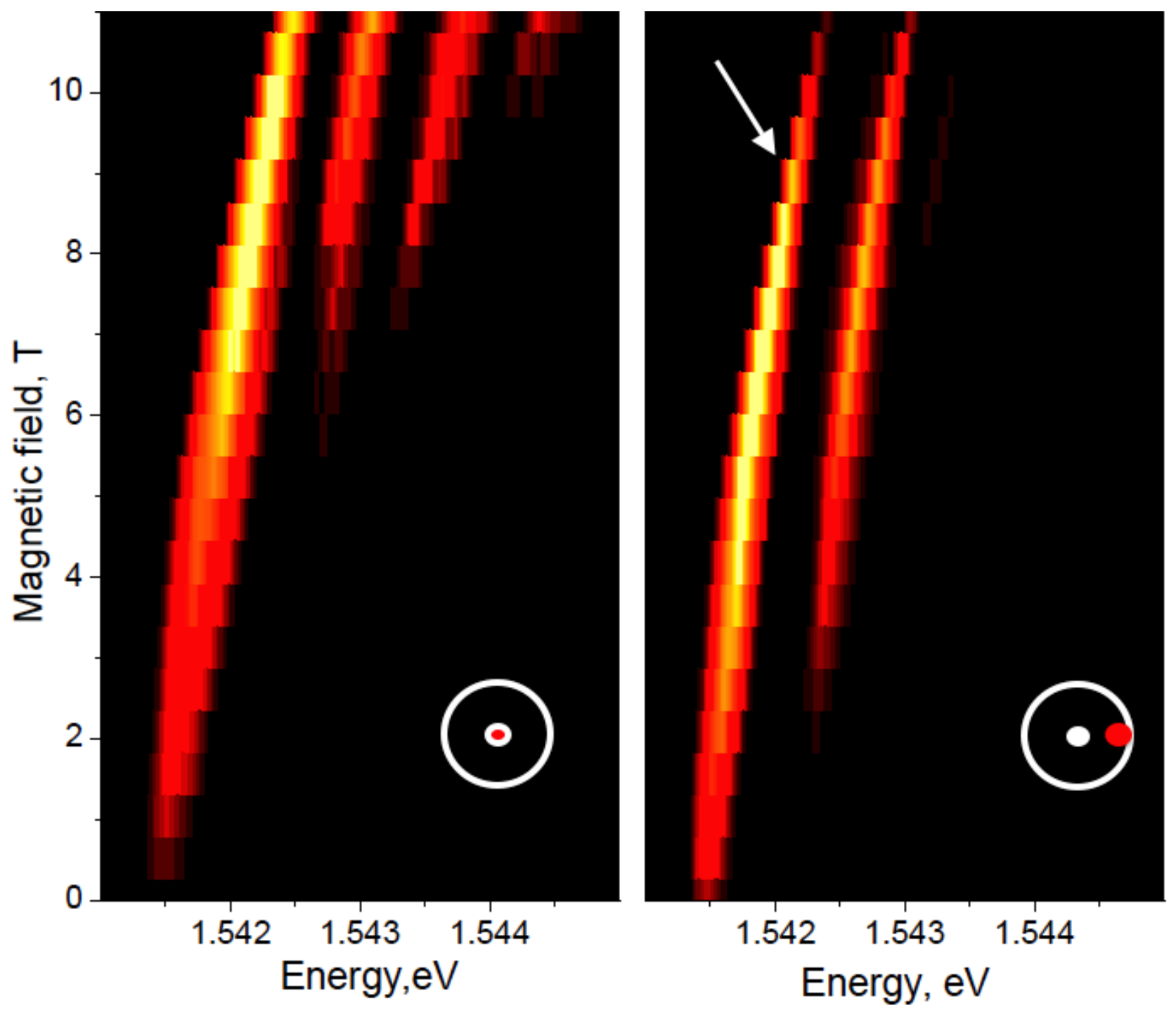

Fig. S 3: **Magneto PL patterns of the 8 $\mu$m diameter micropillar sample in the regime of polariton laser.** The pillar is excited at the center (left panel) and at the edge (right panel), while the signal is detected from the center in both cases.

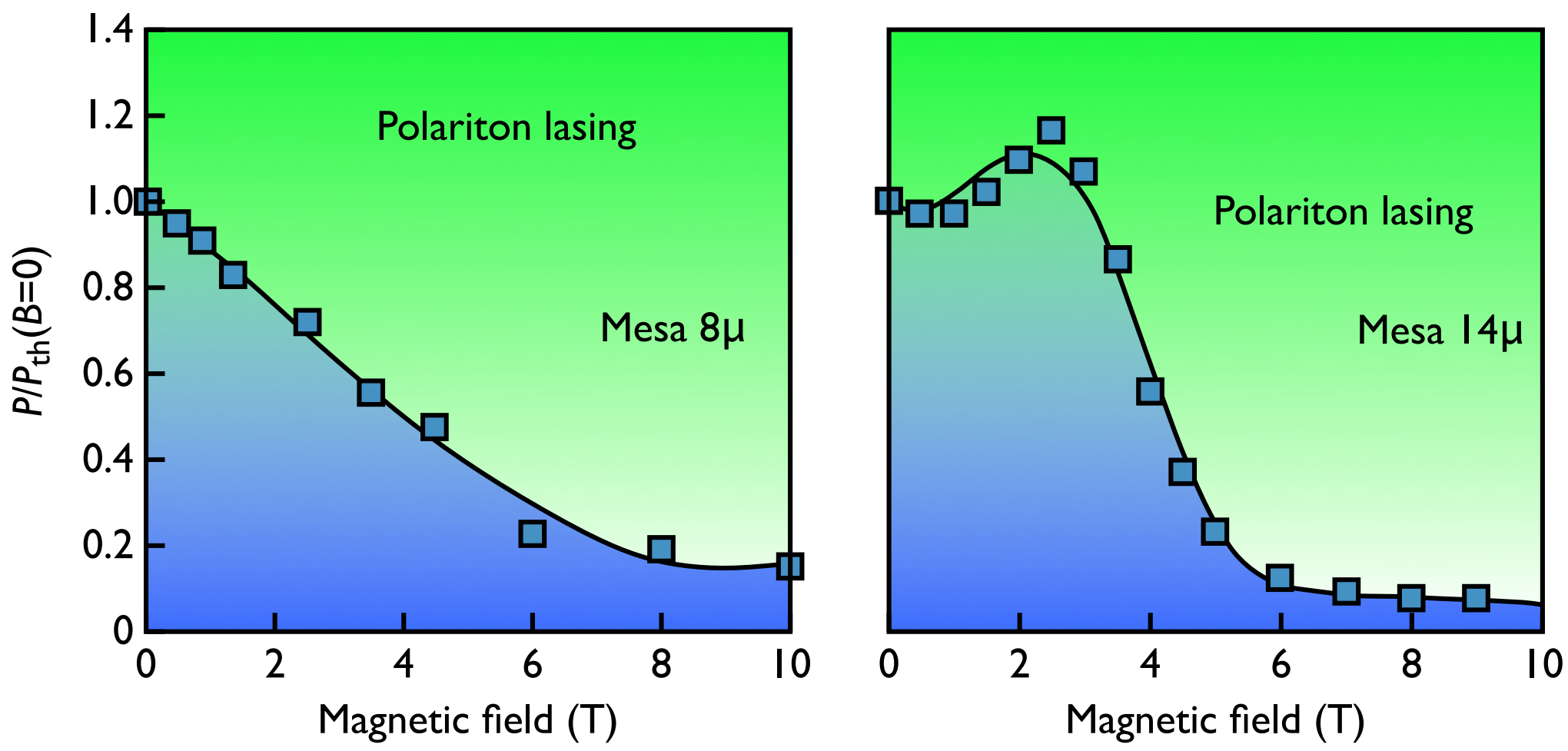

Fig. S 4: **Experimentally observed phase diagrams for polariton emission of the micropillar samples.** Left panel, pillar diameter is 8 $\mu$m, right panel, pillar diameter is 14 $\mu$m.

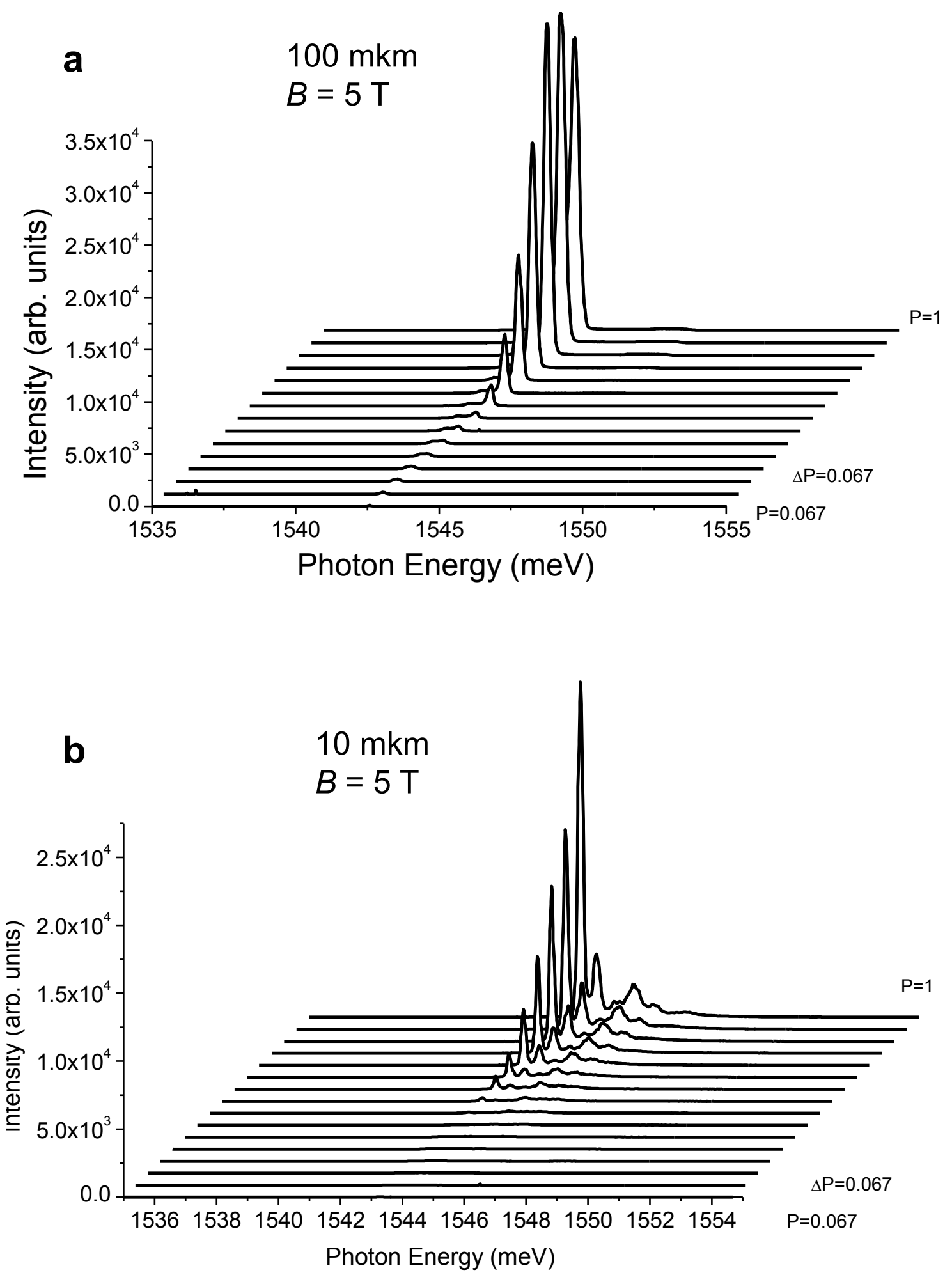

Fig. S 5: **PL spectra of the planar microcavity illustrating the onset of a polariton lasing.** a, Excitation spot diameter is 100 $\mu$m, b, excitation spot diameter is 10 $\mu$m.

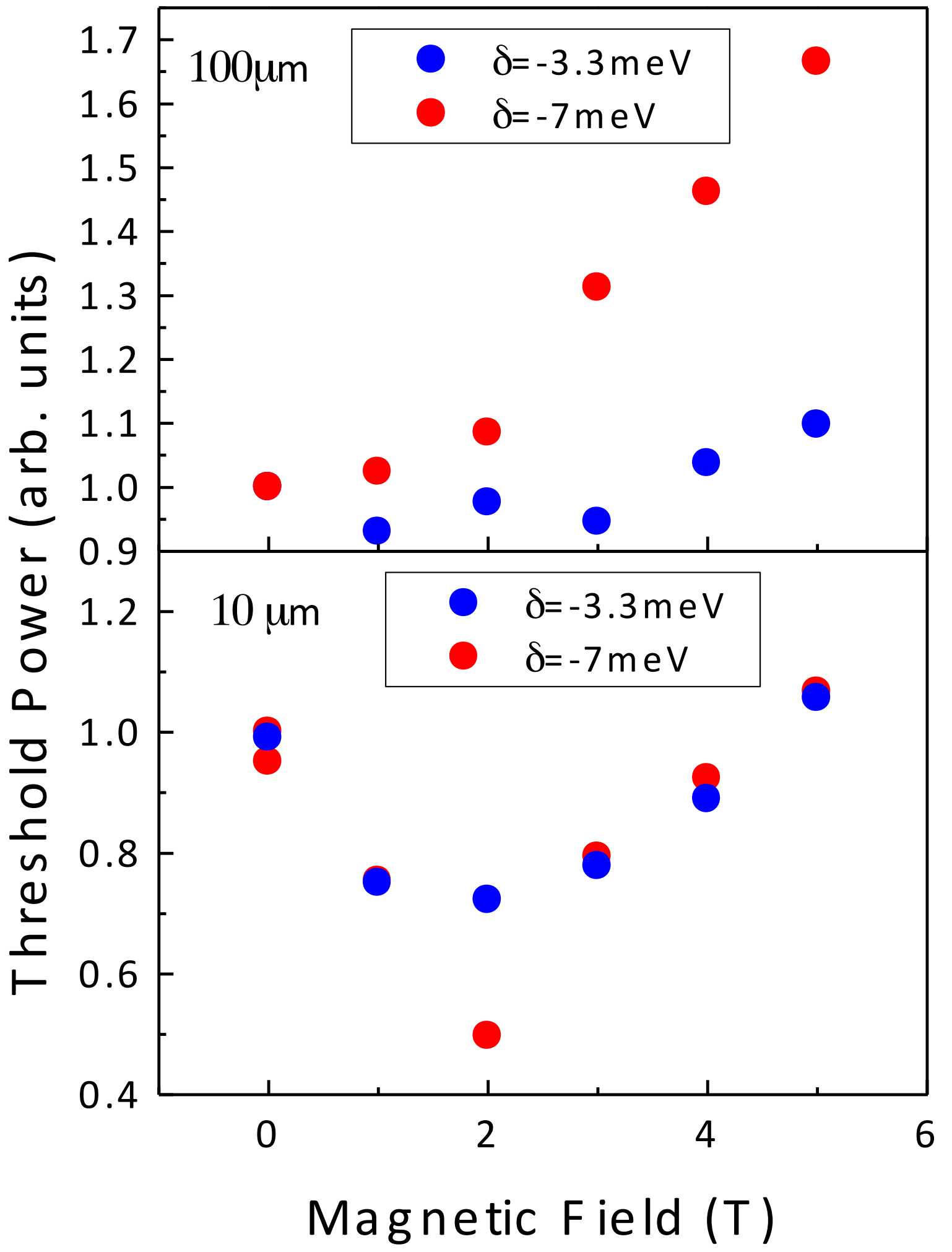

Fig. S 6: **Polariton threshold of the planar microcavity sample.** The data for two different diameters of the excitation spot and two values of detuning is presented.

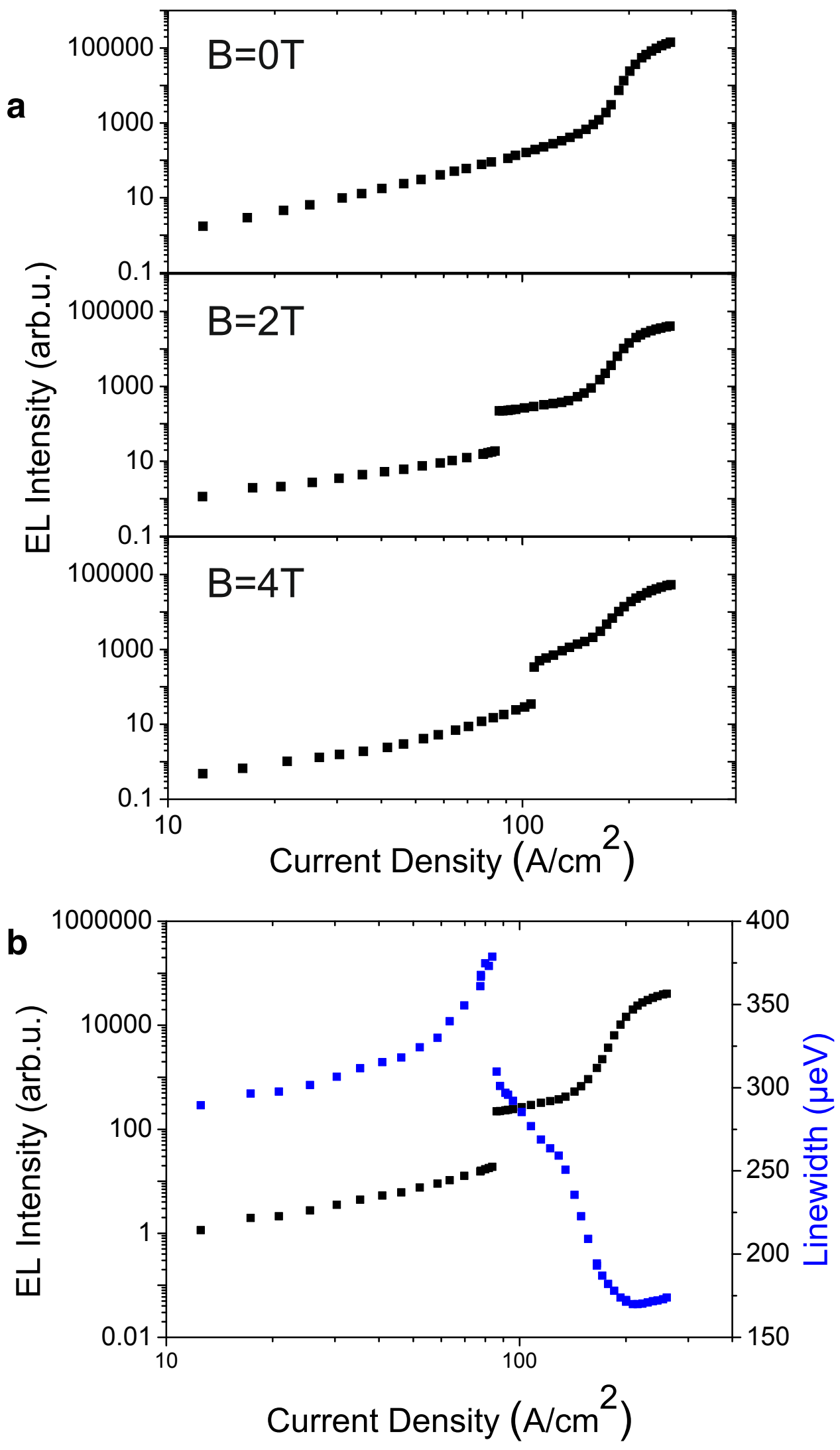

Fig. S 7: **Input-output characteristics and the emission linewidth of the electrically driven polariton laser.** a, Input-output characteristics at a magnetic field of 0 T, 2 T and 4 T. b, Input-output characteristics and the emission linewidth at 2 T.